# Testing multilayer-coated polarizing mirrors for the LAMP soft X-ray telescope

D. Spiga[1§], B. Salmaso[1], R. She[3], K. Tayabaly[1,7], M. Wen[2], R. Banham[4], E. Costa[5], H. Feng[3], A. Giglia[6], Q. Huang[2], F. Muleri[5], G. Pareschi[1], P. Soffitta[5], G. Tagliaferri[1], G. Valsecchi[4], Z. Wang[2]

[1]INAF / Osservatorio Astronomico di Brera, Via Bianchi 46, 23807 Merate (Italy)
[2]Institute of Precision Optical Engineering, Tongji University, 1239 Siping Road, 200092 Shanghai (China)
[3]Department of Engineering Physics, Tsinghua University, 100084 Beijing (China)
[4]Media-Lario Technologies, via Pascolo, 23842 Bosisio Parini (Italy)
[5]INAF/ Istituto Di Astrofisica e Planetologia Spaziali, Via Fosso del Cavaliere 100, 00133 Roma (Italy)
[6]IOM-CNR laboratorio TASC, S.S. 14 Area Science Park, 34149 Basovizza (Italy)
[7]Politecnico di Milano-Bovisa, Via La Masa 1, 20156 Milano (Italy)

## ABSTRACT

The LAMP (Lightweight Asymmetry and Magnetism Probe) X-ray telescope is a mission concept to measure the polarization of X-ray astronomical sources at 250 eV via imaging mirrors that reflect at incidence angles near the polarization angle, i.e., 45 deg. Hence, it will require the adoption of multilayer coatings with a few nanometers d-spacing in order to enhance the reflectivity. The nickel electroforming technology has already been successfully used to fabricate the high angular resolution imaging mirrors of the X-ray telescopes SAX, XMM-Newton, and Swift/XRT. We are investigating this consolidated technology as a possible technique to manufacture focusing mirrors for LAMP. Although the very good reflectivity performances of this kind of mirrors were already demonstrated in grazing incidence, the reflectivity and the scattering properties have not been tested directly at the unusually large angle of 45 deg. Other possible substrates are represented by thin glass foils or silicon wafers. In this paper we present the results of the X-ray reflectivity campaign performed at the BEAR beamline of Elettra - Sincrotrone Trieste on multilayer coatings of various composition (Cr/C, Co/C), deposited with different sputtering parameters on nickel, silicon, and glass substrates, using polarized X-rays in the spectral range 240 - 290 eV.

**Keywords:** X-ray mirrors, X-ray polarimetry, Brewster angle, multilayer mirrors, LAMP

## 1. INTRODUCTION

Astronomical X-ray polarimetry is, as of today, quite an unexplored field: with the noticeable exceptions of the spatially-averaged Crab nebula and an upper limit of a few percent to Sco X-1 and Cyg X-1, no significant polarization has been so far detected in astronomical X-ray sources. Sensitive polarization observations in X-rays would allow us to discriminate between physical models proposed e.g., for shock acceleration in supernova remnants, relativistic jets in blazars, aspherical accretion in X-ray binaries, or X-ray reflection nebulae to name a few. One possibility to manufacture an imaging polarimetric X-ray observatory consists of endowing a focusing X-ray telescope with a pixelated detector having polarimetric capabilities:[1] this is the concept adopted for XIPE (X-ray Imaging Polarimetry Explorer), proposed to ESA for the M4 call with launch in 2025, and currently selected for study phase.

An alternative approach is, however, the one adopted by the Lightweight Asymmetry and Magnetism Probe (LAMP) project. LAMP is a micro-satellite under study to measure the polarization of X-ray sources at the energy of 250 eV, in which the polarization of the astronomical X-ray targets is analyzed via imaging mirrors reflecting at the polarization angle. The scientific targets and the payload description of LAMP are described in detail in another SPIE paper[3].

In the LAMP telescope (Fig. 1, left), an array of 16 paraboloidal mirrors with radius from 201 to 286 mm is mounted around the optical axis of the telescope, reflecting at incidence angles near 45 deg. X-rays near 250 eV ($\lambda$ = 5 nm) are focused to Ar/CO$_2$ gas imaging detectors at a 120 mm distance: in these conditions, only the component perpendicular to the incidence plane (s-polarization) is reflected, while the parallel component (p-polarization) is absorbed. If the X-ray source is linearly polarized, but the polarization plane is rotated by a $\varphi$ angle with respect to the mirror surface, the reflected intensity is proportional to $\cos^2\varphi$. In the general case of a partially polarized X-ray source, the modulation of the image intensity, as recorded by the detector array, is still modulated with amplitude proportional to the degree of linear polarization.

---

[§] contact author: Daniele Spiga, email: daniele.spiga@brera.inaf.it, phone: +39-02-72320427

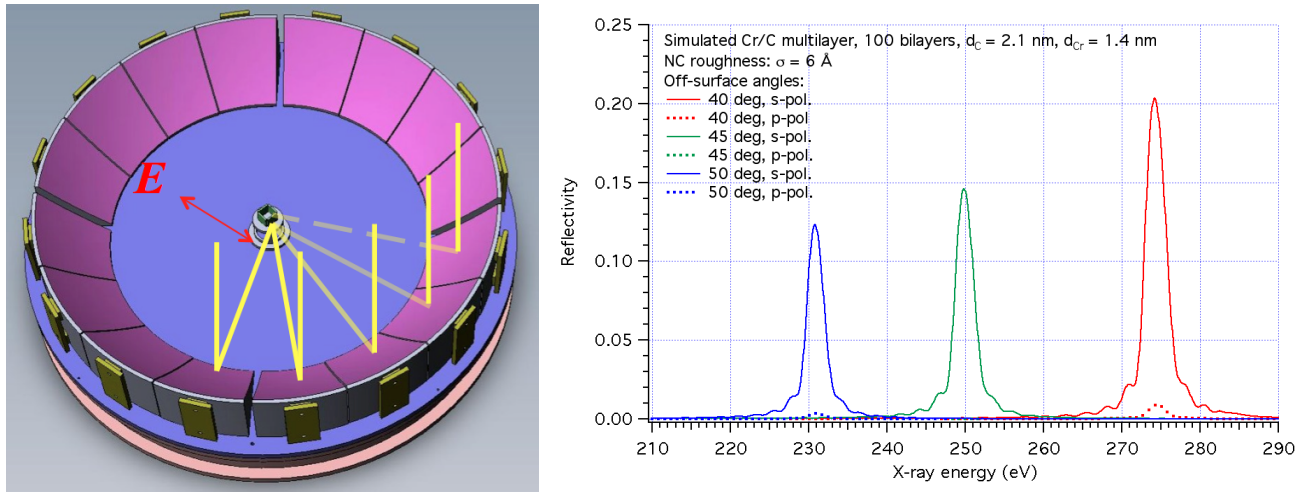

Fig. 1: (left) a layout of the LAMP polarimetric telescope (after[3]). The optical system consists of an array of 16 paraboloidal segments at 45 deg, focusing toward an array of imaging detectors. (right) simulated reflectivity of a polarizing multilayer for s- and p-polarization.

Since in X-rays the refractive index $n$ of all the materials, far from absorption edges, is very close to 1, the off-surface polarization angle $\arctan(1/n)$ is almost exactly 45 deg: in reality, since parabolic mirrors of this size and near this angle exhibit a marked curvature, the mirror longitudinal slope varies from 40 to 50 deg, but the polarizing effect is still present. In fact, the reflectivity of the p-polarization at 40 deg and 50 deg is less than 1% (Fig. 1, right).

Unlike grazing-incidence soft X-ray optics, for which a simple high-Z layer (e.g., gold) is sufficient, multilayer coatings are needed to have high reflectivity at 45 deg incidence. The multilayers have to be periodic to enhance the reflectivity near 250 eV, according to the first-order Bragg law (we do not account for the refractive correction because it is negligible at large angles):

$$2d \sin \theta = \lambda \qquad (1)$$

where $\theta$ is the incidence angle measured off-surface, $d$ is the multilayer d-spacing, and $\lambda$ the X-ray wavelength. Replacing in Eq. 1 the values for $\theta$ and $\lambda$ we obtain $d$ = 3.5 nm. However, the deposition of high-reflectivity multilayer coatings made of several (approx. 100) layer pairs with this low thickness is definitely challenging: not only a tight control on the layer thickness has to be achieved, but also an excellent smoothness of the interfaces has to be maintained throughout the stack. In order to deposit multilayer coatings with these characteristics, a reactive sputtering chamber[4] has been developed at the *Institute of Precision Optical Engineering* (IPOE, Shanghai). To demonstrate the multilayer reflectivity performances, representative Co/C and Cr/C samples have been deposited onto different kinds of substrates (silicon, glass, electroformed nickel), varying the composition of the sputtering gas (pure argon, or argon with a few percent of nitrogen), and therefore the nitruration degree of chromium, cobalt, and carbon. The multilayers can be deposited onto mirror shells in electroformed nickel, the well-experienced technology used to manufacture cheap and high-angular resolution X-ray mirrors, e.g., those of SAX, XMM-Newton, and Swift/XRT, developed at INAF/OAB (*Osservatorio Astronomico di Brera*) and *Media-Lario technologies* (MLT). Since these mirrors do not have to be nested in densely stacked, co-axial systems, lightweight technologies (silicon or glass optics) are not necessary for LAMP and electroformed nickel mirrors represent a natural solution. Moreover, this technology is perfectly suitable also to achieve angular resolutions near 20 arcsec HEW (*Half-Energy-Width*), i.e., better than the required value for LAMP (~1 arcmin HEW).

A direct performance proof requires testing the multilayer coatings at the X-ray energy they are designed for, i.e., 250 eV. This is quite difficult to achieve with commercial equipment, because this energy is heavily absorbed in air. On the other hand, measurements at higher energies (e.g., the widespread Cu-Kα line at 8.045 keV) are difficult to extrapolate at 250 eV in an affordable way. In addition, in order to directly test the polarimetric efficiency of these mirrors, X-rays should be almost completely polarized, and filtered to a passing band narrower than 0.5 eV. These requirements suggest synchrotron radiation as a natural possibility to perform reflectance tests in a fast and efficient way. However, a reflectivity measurement at 45 deg of incidence degrees also requires a precise goniometric system, able to rotate the sample in the incidence plane, and also about the incidence direction in a very large measuring range. Fortunately, the BEAR (*Bending magnet for Emission, Absorption and Reflectivity*) beamline at the Elettra light source (Sincrotrone Trieste, Italy) is fully equipped for reflectivity measurements with these characteristics.

In this paper we show the results of an XRR (X-ray reflectivity) characterization campaign performed at BEAR on multilayer samples for the LAMP telescope. Substrate samples in electroformed nickel have been provided by MLT by replication of a highly polished, 2-inch diameter fused silica master by *General Optics*, with standardized roughness of ~ 1 Å at spatial wavelengths below 1 μm. The masters were coated with a 50 nm thick gold layer deposited by e-beam evaporation, electroformed with nickel, and separated from the master. The thickness of the gold layer is kept to an optimal value[5] to minimize the roughness growth. The resulting disks have been characterized in roughness with the AFM (Atomic Force Microscope) and then diced by electroerosion into squared samples (13 mm x 13 mm, 1 mm thick). The samples were later re-measured with the AFM, to make sure that the electroerosion process did not degrade the surface smoothness. Eventually, the samples were coated at IPOE with Cr/C and Co/C multilayers, with variable percent of nitrogen in the sputtering gas (pure argon, argon + 2%, 4%, and 6% of $N_2$) aiming at a reflectivity enhancement. The nominal multilayer recipe is a stack of 100 couples of layers, where Co (or Cr) layers are 1.4 nm thick and C layers are 2.1 nm thick. During the same coating run, squared samples of silicon wafer and D263 glass were coated to disentangle the roughness introduced by the multilayer growth from the one of the substrate. The resulting samples were tested at the BEAR beamline in polarized X-rays from 240 to 290 eV, in the 40-50 deg angular range. In Sect. 2 we discuss the issue of microroughness for mirrors at 45 deg. In Sect. 3 we describe some measurements performed before the campaign at BEAR, and in Sect. 4 we introduce the experimental setup of the beamline. Measurement data with the relative analysis of the reflectivity curves are shown in Sect. 5 and 6.

## 2. THE ROUGHNESS ISSUE

The limit to the performance of an X-ray mirror is the roughness of the interfaces, including those between consecutive layers in a multilayer. In grazing-incidence X-ray mirrors, a 250 eV energy is low enough to minimize the impact of the roughness, but at 45 deg the situation is different. This is also true for the interdiffusion of layers, which have an identical effect on the specular reflectivity. In fact, an additional X-ray scattering (XRS) measurement is needed to discriminate the roughness from interdiffusion, while this is not possible by a simple XRR measurement. In this paper we will consider the combined effect of the roughness and the interdiffusion into the same σ parameter, representing the interface profile rms.

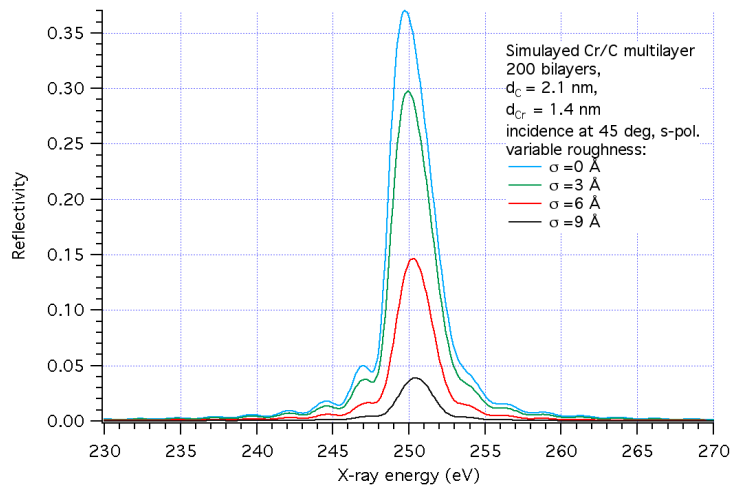

Fig. 2: simulated multilayer reflectivity at the polarization angle, for variable roughness values and in s-polarized X-rays. In order to preserve the peak reflectivity, the surface roughness has to be much smaller than the d-spacing (Eq. 3).

The reflectivity of a surface decreases exponentially with the square of σ sin$\theta$/$\lambda$ (Debye-Waller formula):

$$R_\sigma = R_0 e^{-\left(\frac{4\pi\sigma \sin\theta}{\lambda}\right)^2}. \quad (2)$$

In grazing incidence X-ray mirrors, $\theta$ is shallow and contributes to mitigate the roughness impact. However, at 45 deg, the impact on the reflectivity is much more severe. To put it another way, using Eq. 1, the previous equation becomes

$$R_\sigma = R_0 e^{-\left(\frac{2\pi\sigma}{d}\right)^2}, \quad (3)$$

and we see that the peak reflectivity decreases exponentially with the square of the $\sigma$/$d$ ratio. Therefore, the deposition of high reflectivity multilayer requires an interface rms much smaller than the d-spacing. Since $d$ = 3.5 nm, the roughness has to be less than a few angstrom, exactly as required in hard X-ray optics. In Fig. 2 we display the expected Bragg peak of a Cr/C multilayer, assuming different values of $\sigma$.

The definition of the $\sigma$ parameter, however, should be referred to a window of spatial wavelengths. If the reflectivity reduction is caused by roughness, the inferred value of $\sigma$ depends on the angular acceptance $\Delta\theta$ of the detector, i.e., on the amount of scattered radiation that is collected by the detector window. Combining Eq. 1 with the grating formula in the limit of small $\Delta\theta$, we find that the spatial wavelengths that are excluded and contribute to the measured roughness are

$$l < \frac{2\lambda}{\sin\theta\ \Delta\theta} = \frac{4d \sin\theta}{\sin\theta\ \Delta\theta} = \frac{4d}{\Delta\theta} \quad : \quad (4)$$

in the XRR setup used at BEAR, $\Delta\theta$ = 2.8 deg, i.e., $l$ < 0.3 μm. Therefore, we may expect the result to approximately fit an $\sigma$ value measured over a scan range of a 1 μm. Roughness measurements on coated nickel samples before and after coating are shown in the next section. In contrast, in presence of layer interdiffusion the multilayer transmittance is increased; hence, the corresponding reflectivity reduction cannot be compensated enlarging $\Delta\theta$, and the XRR scan fits a $\sigma$ value higher than the one derived from surface roughness metrology.

## 3. PRELIMINARY SAMPLE ANALYSIS

The roughness of the nickel samples was measured with the AFM in INAF/OAB before being shipped to IPOE to make sure that the roughness is within the limits specified in the previous section. After the coating, the roughness was re-measured to detect possible roughness degradations introduced by the coating process.[6] Some measured maps are shown in Fig. 3: typical values are around 6 Å rms over a 10 μm scan and 3.5 Å over a 2 μm scan. Roughness measurements were repeated, after coating, over the 10 μm range (usually the most concerned one by the issue of roughness growth).[7] We did not detect any relevant change in the roughness morphology. Neither we did observe a significant increase of the rms value. This rules roughness amplification out in the stack.

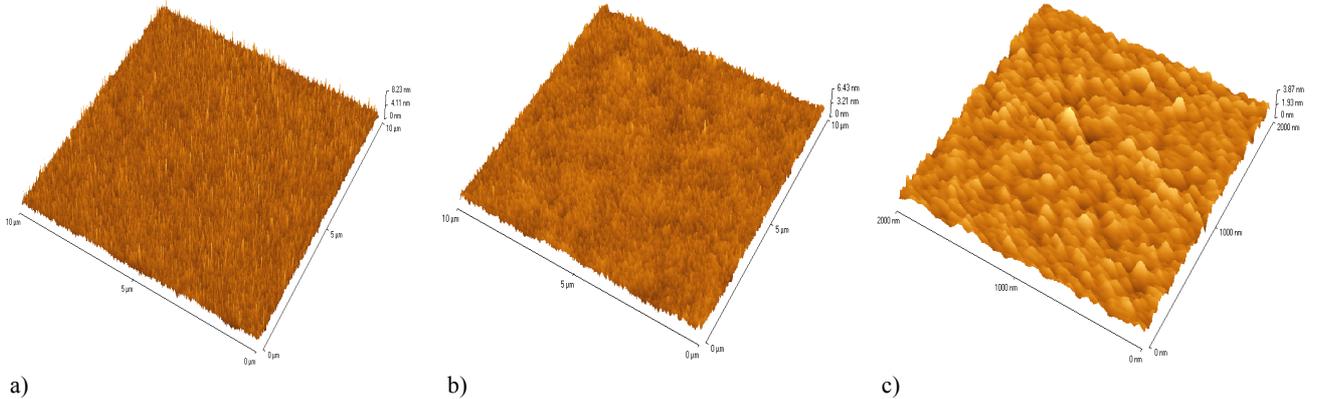

a) b) c)

Fig. 3: AFM measurements of nickel samples: a) before coating: 10 μm scan, σ = 6.2 Å; b) after coating, 10 μm scan, σ = 5.4 Å; c) before coating, 2 μm scan, σ = 3.4 Å.

Even though the witness samples with silicon or glass substrates were not measured with the AFM, we can assume the initial silicon roughness to be 0.7 Å rms, i.e., typical of the pristine surface of commercial silicon wafers,[8] and that it is kept unchanged after the coating deposition. The same assumption suggests a value of 2 Å for the roughness of the D263 glass substrates, as known from an extrapolation of measurements performed at INAF/OAB on D263 samples.

Another test on coated samples was performed using an X-ray diffractometer at the standard Cu-Kα fluorescence line (8.045 keV, 73% polarized by the crystal monochromators). Grazing-incidence XRR scans in theta-2theta geometry were acquired at IPOE and modeled using the IMD program:[9] some measured scans with the respective modeling are shown in Fig. 4. This kind of test returns an in-depth characterization, a measurement of the multilayer reflectivity performance, and an independent estimation of the interface rms in the stack. The peaks appear sharp and intense, denoting a good periodicity of the multilayer stack and good interface smoothness. However, the inferred rms of the interface ($\sigma$ = 3.5 Å for multilayer on silicon, $\sigma$ = 5.5 Å for multilayers on nickel) is higher than the one measured with AFM measurements over 2 μm (see Fig. 3). In addition, the angular scale is not sufficiently resolved to assess the thickness uniformity in the stack. We will see in the next sections that the measurement at BEAR will allow us to detect asymmetries in the peak profile, from which we can retrieve information on the stack regularity.

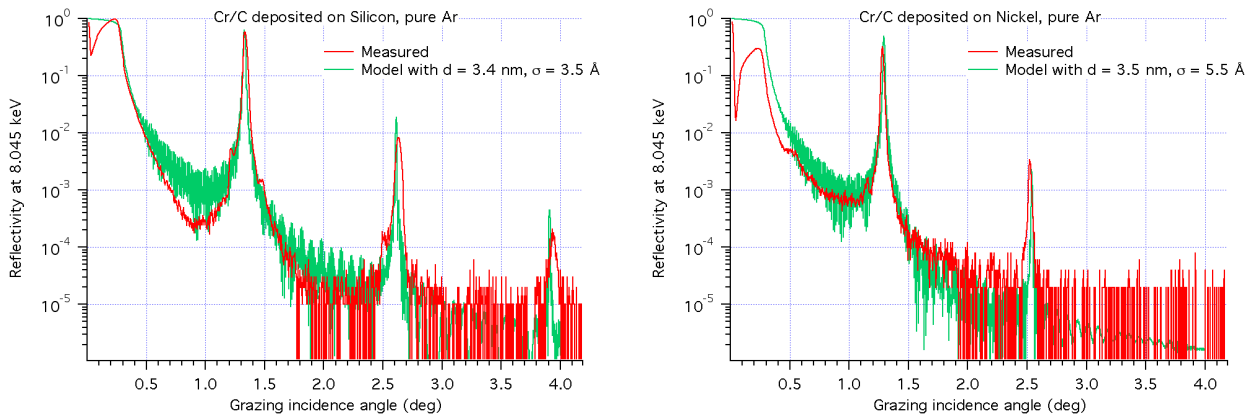

Fig. 4: sample reflectivities measured at 8.045 keV; the measurements are compared to reflectivity models, assuming perfectly periodic multilayer structures.

## 4. EXPERIMENTAL SETUP

BEAR is a bending magnet beamline at Elettra (Sincrotrone Trieste) dedicated, among other things, to reflectivity measurements.[10],[11] BEAR covers a band of light wavelengths from visible light to 1.6 keV, delivering radiation with well-defined polarization and spectral purity properties. The sample chamber is equipped with motors enabling movements of the sample holder in all directions, accurate alignment, and reflectivity scans in a very wide angular range. Finally, a sample insertion chamber is available to change a sample without venting the chamber, thereby enabling the measurement of a large number of samples in the time allocated for the experiment. For these reasons, BEAR was selected to perform the reflectivity characterization of the LAMP samples. A layout of the beamline is shown in Fig. 5.

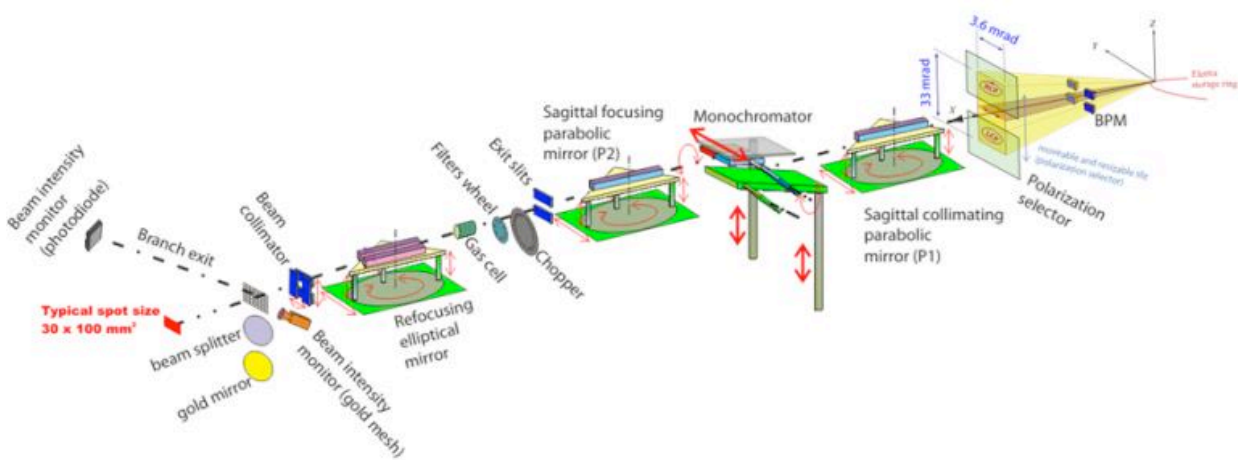

Fig. 5: layout of the BEAR beamline at Elettra – Sincrotrone Trieste. At the entrance of the beamline, the polarization selector allows selecting linearly polarized light (http://www.elettra.trieste.it/lightsources/elettra/elettra-beamlines/bear/beamline-description.html).

The bending magnet radiation is purely linearly polarized in the plane of the electron orbit. When viewing the source from out the orbit plane, a small elliptical component appears, corresponding to the projection of the electron orbit on the observation plane. The amount of linear polarization depends on the X-ray energy and the acceptance angle of the slit at the front end of the beamline (the polarization selector in Fig. 5). For this experiment, we needed a linearly polarized beam; hence, a 2 mm aperture was selected, corresponding to a 98% of linear polarization. The remaining 2% of circular polarization only conveys a 1% of polarization in the vertical plane, yielding 99% of polarized rays in the horizontal plane.

After passing through the selector, the beam is collimated in the sagittal direction by a parabolic mirror and spectrally filtered in the soft X-ray band by the grazing incidence monochromator G1200. For a fixed deviation angle, the degree of monochromation depends on the amplitude of the vertical amplitude of the slit. We have selected a highly monochromatic setup: a 70 μm wide aperture corresponding to an energy resolution $E/\Delta E$ = 1500, i.e. $\Delta E$ = 0.16 eV at $E$ = 250 eV. The selected energy can be scanned varying the incidence angle on the grating. After being refocused to the exit slits by another parabolic mirror, the beam traverses a 0.6 μm Ag filter that removes higher order harmonics.

Finally, the X-ray beam is refocused to the sample via an elliptical mirror, enabling spatially resolved measurements at different locations on the sample holder. The beam location can be seen setting the monochromator to zero order and removing the filter. This way, also the visible component passes through the monochromation chain and the illuminated location appears as a bright spot on the sample, which can be viewed by one of the cameras located in the manipulator chamber. The focused beam converges within a cone of 0.06 deg of amplitude, resulting in an equal divergence after reflection. The photocurrent of the elliptical mirror is used to monitor the incoming beam intensity.

The sample holder can be translated in all directions, aligned in tilt angle and rotated in the incidence plane, in a wide range of $\theta$ angles (1 to 87 deg off-surface). Even if up to eight samples at a time could be mounted on the sample holder, each sample was aligned individually. This is achieved to a 0.1 deg accuracy using photodetectors at fixed, symmetric positions with respect to the incident beam. When the sample is aligned and set at the incidence angle of 45 deg, the monochromatic setup is restored and the detector is aligned, scanning and centering the reflected beam. The detector used is a photodiode located 160 mm from the sample, with an entrance window of 8 mm × 8 mm, yielding an acceptance angle of $\Delta\theta$ = 2.8 deg. The large value of $\Delta\theta$ allows us including a very large fraction of the scattered beam; hence, following the reasoning reported in Sect. 2 (Eq. 4) the reflectivity measurement should fit surface roughness measurements over a 1 μm range (i.e., the rms measured by an AFM over a region of 1 μm size). The divergence caused by the beam refocusing is much smaller than $\Delta\theta$ and can be neglected in the spatial wavelength computation.

Finally, the polarization properties of the samples are measured rotating the entire sample chamber (Fig. 6) around the beam axis by the angle $\varphi$ from 0 deg (vertical incidence plane) to 90 deg (horizontal incidence plane).

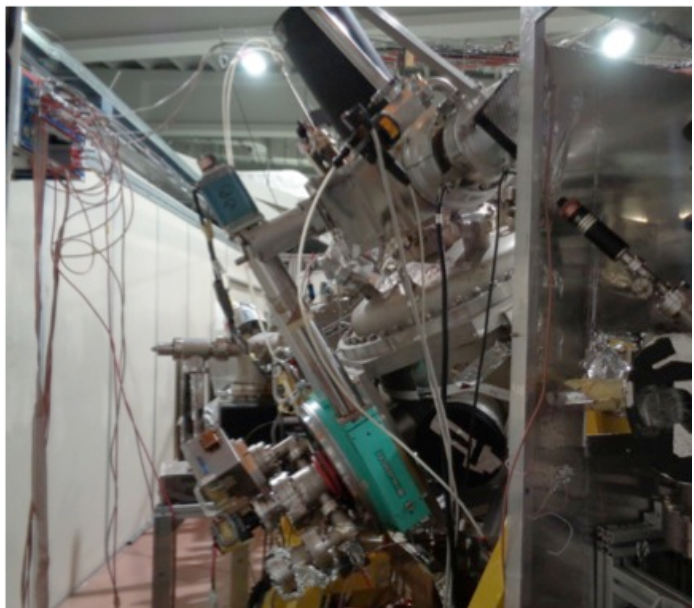

Fig. 6: rotation of the experimental chamber of the BEAR beamline about the incident beam to change the orientation of the incidence plane with respect to the polarization plane.

## 5. EXPERIMENTAL RESULTS

For every sample, two kinds of measurement were done: angular scans and energy scans. Angular scans are performed at a fixed setting of the monochromator, recording the current $I_R$ induced in the detector photodiode by the reflected beam while scanning the $\theta$ angle. Energy scans are obtained keeping the sample in a fixed position and recording the photodiode signal $I_R$ while scanning the energy $E$. During the scan, the beam monitor records a signal $I_{RM}$ proportional to the incident flux intensity, in order to correct possible flux fluctuations in the source. Each scan is followed by a "dark" background count $I_{RB}$. The direct beam, $I_D$, was previously recorded along with its dark count $I_{DB}$ and its monitor signal $I_{DM}$. The reflectivity, as a function of either $\theta$ or $E$, is computed using the following equation:

$$R(\theta, E) = \frac{\frac{I_R(\theta, E) - I_{RB}}{I_{RM}(\theta, E)}}{\frac{I_D(\theta, E) - I_{DB}}{I_{DM}(\theta, E)}} \tag{5}$$

Some angular XRR scans are displayed in Fig. 7, with the incidence plane perpendicular to the main polarization plane (s-polarization, $\varphi = 0$ deg). We displayed on the left side the angular scans at 250 eV of the samples deposited on silicon substrates, deposited with variable percent of $N_2$ in the sputtering gas. The XRR exhibits a clear Bragg peak in the vicinities of the 45 deg polarization angle. The peak position is spread within a 3 deg range, although still in the range of incidence angles for LAMP (40-50 deg). Translated in terms of d-spacing fluctuation, this angular spread amounts to a maximum d-spacing error of 2 Å. Such an error might be caused by the different position of the samples in the coating chamber. On the right side of Fig. 7, we show the angular scans at different energies, proving that the XRR peak correctly moves toward shallower angles as the X-ray energy is increased.

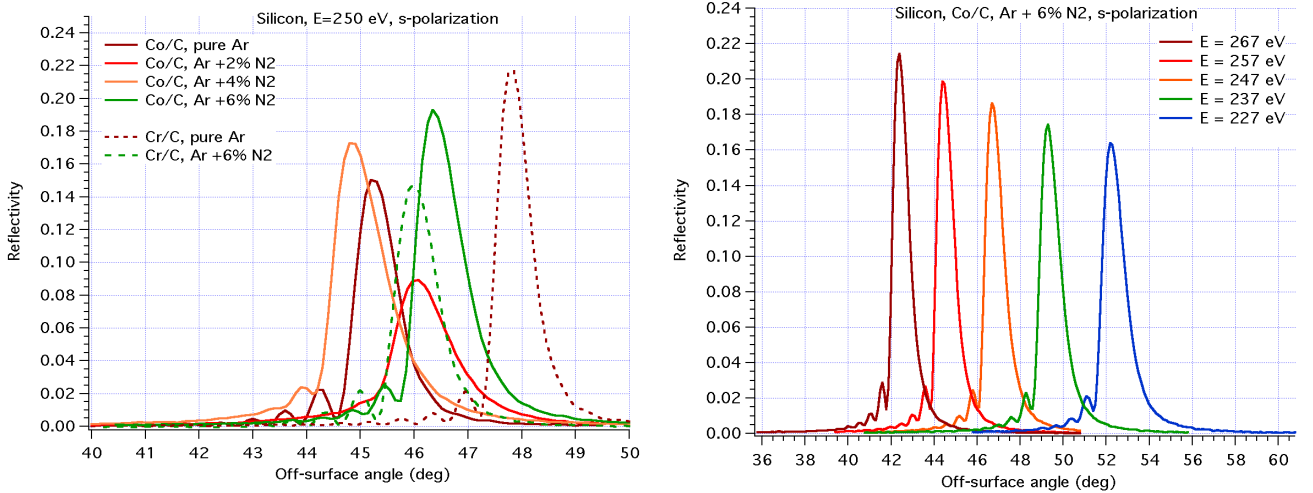

Fig. 7: angular (theta-2theta) reflectivity scans of a few samples deposited on Silicon: (left) at the fixed X-ray energy of 250 eV and (right) a single sample at different monochromatic energies.

Inspection of the reflectivity curves in Fig. 7 shows that the peaks are not exactly symmetric: in fact, they exhibit pronounced secondary peaks on the side of shallower incidence angles. This suggests some a-periodicity of the stack, even if it cannot be immediately evaluated. We will see in Sect. 6 that an accurate fit of the XRR angular scan allows us a more precise determination of the thickness drift of Co, Cr, and C throughout the stack.

The Bragg peaks are also characterized by different heights, depending on the amount of nitrogen in the gas used for the sputtering process. Apparently, the reflectivity of Co/C increases with the nitrogen inlet (at least, up to 6%), keeping the interfaces smooth and abrupt during the multilayer growth. On the contrary, reactive sputtering with nitrogen degrades the performances of Cr/C multilayers: in fact, the maximum reflectivity (21.7%) at 250 eV is exactly found with the Cr/C multilayer deposited without pure argon.

Angular XRR scans for multilayers deposited on glass samples and nickel substrates are shown in Fig. 8. The reflectivity of these multilayer stacks is lower than for samples deposited on silicon wafers, as expected from the different substrate roughness (Sect. 3), but the Bragg peak is still well pronounced. For nickel samples, the Bragg peak is found at larger angles than those of the other samples: this means that the d-spacing is smaller, probably because of the different position in the sputtering chamber. The dispersion of the d-spacing and the nitruration effect on the reflectivity of the multilayers exhibit a behavior similar to the one shown in Fig. 7. Also energy scans on nickel samples (Fig. 9) clearly show that the presence of nitrogen is beneficial for Co/C multilayers and harmful for Cr/C multilayers. If we consider only the multilayers deposited in the best conditions (Cr/C deposited without nitrogen, Co/C deposited with nitrogen), and excluding a roughness growth in the stack (as suggested by AFM measurements, see Fig. 3), the peak reflectivity is consistent with an interface rms of 3.5 - 4 Å for silicon samples, 4.5 Å for glass samples, 6.5 Å for nickel samples. These numbers are 2.5 - 3 Å higher than the roughness values measured by the AFM over scales of a few microns (Sect. 3). The difference might be ascribed to interlayer diffusion.

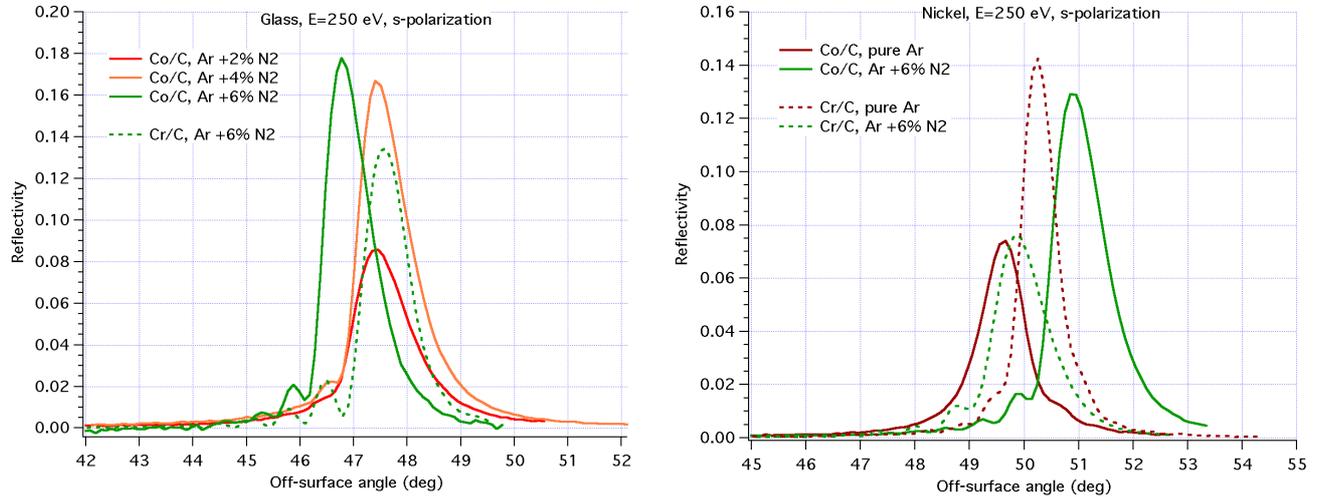

Fig. 8: angular reflectivity scans of a few samples deposited on D263 glass (left) and electroformed nickel (right), at the fixed X-ray energy of 250 eV.

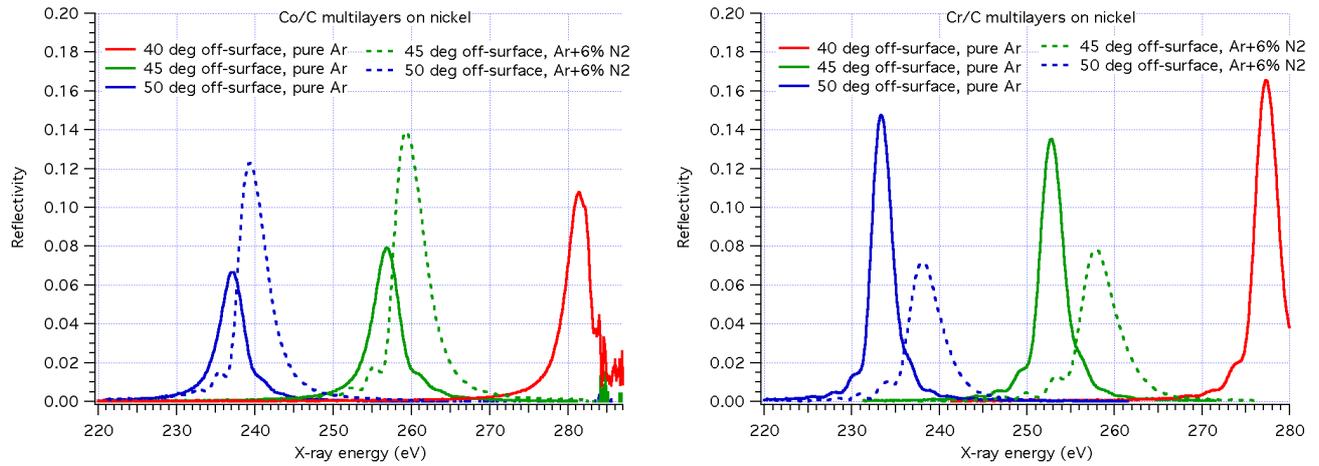

Fig. 9: energy scans at fixed incidence angles, for some Co/C multilayers (left) and Cr/C multilayers (right) tested in this measurement campaign at BEAR. Because of the absorption in the stack, the measurements could not be extended beyond the K-edge of carbon.

Reflectivity measurements were also acquired in p-polarization, i.e., after rotating the experimental chamber (Fig. 6) to $\varphi = 90$ deg, always keeping the incidence near the polarization angle. In these conditions, as expected, the reflectivity is much less then 1% and mostly related to the residual elliptical polarization in the incident beam. Some peak reflectivity values are reported in Tab. 1.

The last kind of measurement performed was aimed at investigating the polarization performances of a few selected samples. XRR angular scans at 250 eV were iterated for ten different orientations of the $\varphi$ angle, from 0 to 90 deg. The resulting reflectivity curves are shown in Fig. 10, left: the angular scales of the different scans are not exactly aligned because the $\theta$ alignment is not exactly maintained during the rotation in $\varphi$. Fig. 10 (right) also shows that the peak reflectivity decreases from nearly 20% to 0.2% as the incidence plane changes from almost purely s-polarization to almost purely p-polarization, following the well-known $\cos^2\varphi$ law of polarizers. This is the behavior expected from the array of 16 mirrors of LAMP (Fig. 1) in presence of an almost perfect polarization. As already noticed, at $\varphi = 90$ deg the measured reflectivity is 0.2%, instead of being near the detector background level (approximately estimated as 0.02%).

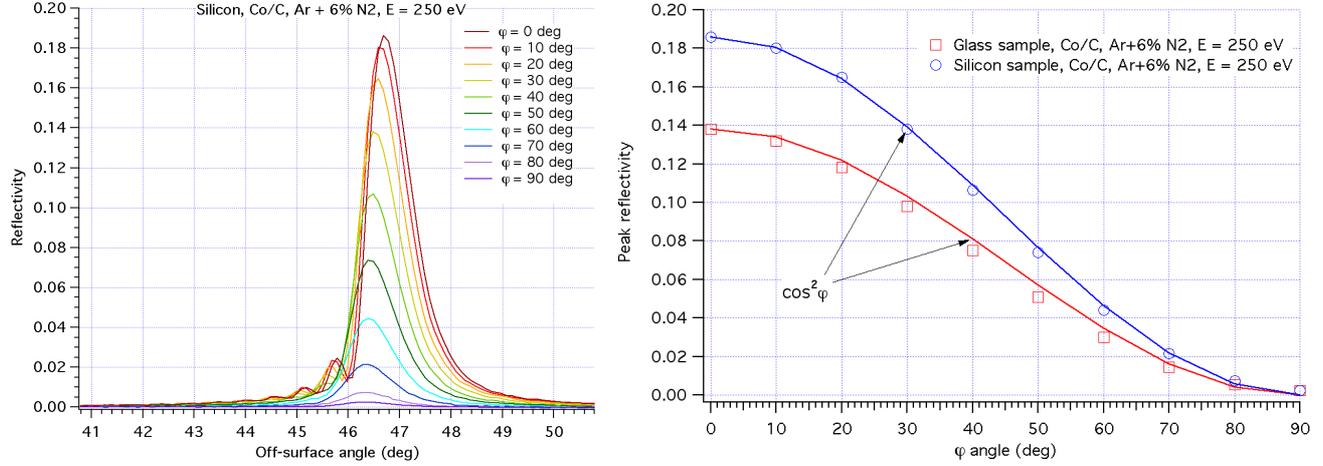

Fig. 10: (left) a set of theta-2theta scans at fixed energy of 250 eV, for different values of the $\varphi$ angle between the polarization plane and the incidence plane, from almost pure s-polarization to almost pure p-polarization; (right) peak reflectivity as a function of the $\varphi$ angle for two samples; the reflectivity perfectly follows the Malus law of polarizers.

However, the presence of a 1% of vertical polarization, passing through the finite aperture of the polarization selector, partly explains this measured reflectivity at $\theta = 45$ deg, $\varphi = 90$ deg. The measured reflectivity at the generic rotation angle $\varphi$ is a combination of the s- and p-component:

$$R_\varphi = R_S(I_0 \cos^2\varphi + I_{90} \sin^2\varphi) + R_P(I_0 \sin^2\varphi + I_{90} \cos^2\varphi) \qquad (6)$$

where $I_0 = 99\%$ and $I_{90} = 1\%$. We can therefore retrieve the s- and p-reflectivity from the measured $R_0$ and $R_{90}$:

$$R_S = \frac{R_0 I_0 - R_{90} I_{90}}{I_0^2 - I_{90}^2} \qquad R_P = \frac{R_{90} I_0 - R_0 I_{90}}{I_0^2 - I_{90}^2} < R_{90} \qquad (7)$$

the computed reflectivities for the two polarizations are reported in Tab. 1.

Tab. 1: peak reflectivity for some representative measurements. We also report the values of the reflectivity values for s- and p-polarization, computed using Eq. 7. The residual value of $R_P$ is close to the detector background and partly caused by Bragg peaks not exactly aligned to the polarization angle.

| Sample | $R_0$ | $R_{90}$ | $R_S$ | $R_P$ |
|---|---|---|---|---|
| 1) Silicon, Co/C, pure Ar | 15% | 0.2% | 15.2% | 0.05% |
| 2) Nickel, Co/C, pure Ar | 7.4% | 0.08% | 7.5% | 0.05% |
| 3) Silicon, Cr/C, pure Ar | 21.7% | N/A | - | - |
| 4) Nickel, Cr/C, pure Ar | 14.5% | N/A | - | - |
| 5) Silicon, Co/C, Ar + 6% $N_2$ | 19.3% | 0.24% | 19.5% | 0.05% |
| 6) Glass, Co/C, Ar + 6% $N_2$ | 17.9% | 0.25% | 18.1% | 0.07% |
| 7) Nickel, Co/C, Ar + 6% $N_2$ | 12.8% | 0.23% | 12.9% | 0.10% |
| 8) Silicon, Cr/C, Ar + 6% $N_2$ | 14.8% | 0.2% | 14.9% | 0.05% |
| 9) Glass, Cr/C, Ar + 6% $N_2$ | 13.4% | 0.28% | 13.5% | 0.15% |
| 10) Nickel, Cr/C, Ar + 6% $N_2$ | 7.6% | 0.12% | 7.7% | 0.04% |

We note that the reflectivity for the p-polarization is now very close to the estimated instrumental noise level, as expected, with the exception of some samples. The highest $R_P$ value is the one of the sample no. 9 (on glass). Some nickel samples (the no. 2) exhibit the same $R_P$ value of the corresponding ones on silicon (or lower, no. 10). Some other (such as the no. 7) anomalously high value of $R_P$ can be related to the Bragg angle of some nickel samples, which is closer to 50 deg than to 45 deg (Fig. 8). We therefore conclude that there is no significant de-polarization effect by the different substrate roughness.

## 6. MULTILAYER STACK ANALYSIS

The interpretation of the reflectivity curves measured at BEAR is made difficult by the slight a-periodicity in the stacks, as witnessed by the secondary peaks on the lower angle side. In fact, a multi-parametric fit is necessary to manage at the same time layer density, thickness, and roughness, throughout a stack of 200 thin layers. To do this, we have used PPM[12] (*Pythonic Program for Multilayers*), a program able to accurately fit XRR curves, thereby extracting the multilayer structure in a very affordable way.[13]

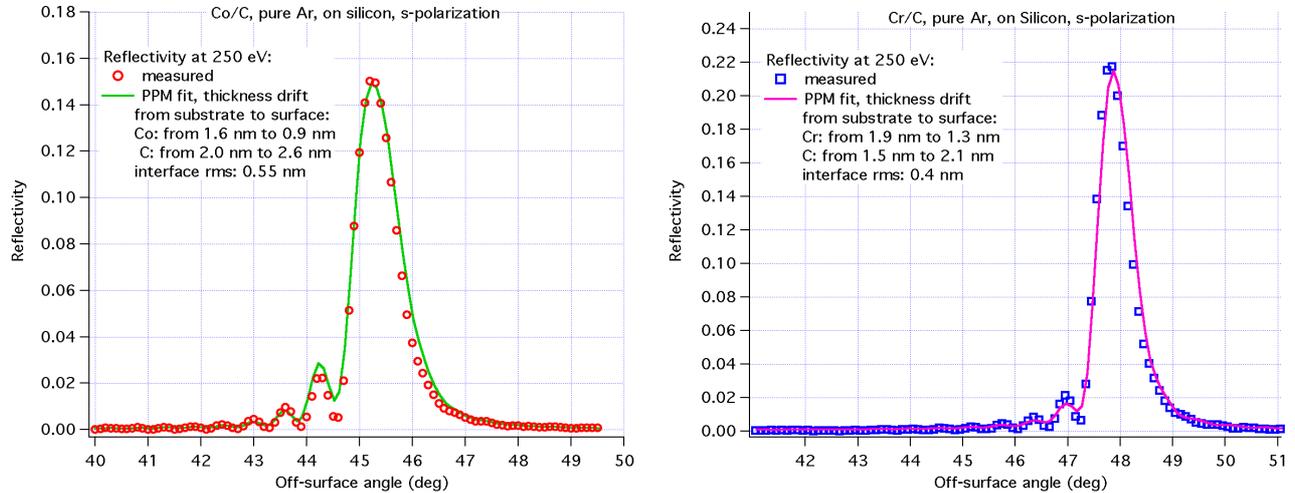

Fig. 11: theta-2theta XRR scans of (left) a Co/C multilayer sample and (right) a Cr/C sample. We also show the reflectivity fit found by the PPM program: the pronounced secondary peaks are interpreted as the effect of a layer thickness drift in the stack.

The PPM fit has been operated on the XRR scans of two samples, Co/C and Cr/C deposited on silicon. We did not use multilayers deposited with nitrogen because the optical constants of layers with compound composition are unknown. We have selected the multilayers deposited on silicon because they are the ones with the highest reflectivity and therefore the clearest reflectance fringes. To avoid increasing the fit complexity, we have assumed constant layer densities in the stack, a second-order polynomial trend of the thickness values, and a linear variation of the interface rms. We therefore adopted as fit parameters: the layer density (constant in the stack), the layer thickness at the surface, the first and the second derivative of the thickness trend, the roughness at the substrate and its derivative. This parameter set is applied to both Co (or Cr) and C layers, assumed to be completely independent of each other. Starting from an initial guess of a perfectly periodic multilayer, with bulk layer density, and roughness equal to the one measured with the AFM (Sect. 3), PPM finds a very satisfactory solution (Fig. 11) for both Co/C and Cr/C multilayer. The results are the following:

- for the Co/C multilayer, the layer thickness of Co decreases (non-linearly) from 16 Å to 9 Å from the substrate toward the surface. The carbon layers, in contrast, increase from 20 Å to 26 Å in thickness. The Cr/C multilayer exhibits a similar behavior: Cr layers are diminished from 19 to 13 Å from substrate to surface, while C layers grow in thickness from 15 Å to 21 Å.
- the layer density of Co and Cr is lower than the respective bulk values (7.0 $g/cm^3$ vs. 8.9 $g/cm^3$ for cobalt, 6.5 $g/cm^3$ vs. 7.1 $g/cm^3$ for chromium). The density of C is also lower (1.6 $g/cm^3$ vs. 2.3 $g/cm^3$ for graphite). A density lower than the natural value is commonly encountered in thin film coatings.
- the interface (including roughness and diffuseness) rms does not increase significantly in the stack. The found values are 5.5 Å for Co/C and 4 Å for Cr/C. As already mentioned, this is higher than the roughness measured with the AFM over 2 μm scans. The difference might be explained by a few angstrom of layer interdiffusion.

## 7. CONCLUSIONS

Reflectivity measurements of polarizing mirror samples with periodic Co/C and Cr/C multilayer coatings, performed in synchrotron light at 250 eV, show that the LAMP approach to astronomical X-ray polarimetry is feasible, even if the unusually large incidence angle (45 deg) poses a challenge to roughness and d-spacing control. Nevertheless, good polarizing and reflectivity properties could be demonstrated also with multilayers deposited on electroformed nickel, a standard material to fabricate the optics of X-ray telescopes. Moreover, the gas composition in the sputtering process shows clear effects: probably, the presence of nitrogen in the sputtering gas reduces the Co and C layer interdiffusion. However, for Cr/C multilayers the effect of nitrogen is detrimental, and pure argon should be used. Finally, the stack structure could be extracted from a detailed fit of X-ray reflectivity curves, therefore assessing the stability of the deposition rate in the sputtering process and the layer interface abruptness.

## ACKNOWLEDGMENTS

We thank Elettra - Sincrotrone Trieste for granting us the beam time at the BEAR beamline and for providing partial financial support (proposal No. 20145169).